\documentclass[aps,prb,reprint,superscriptaddress]{revtex4-1}
\usepackage{graphicx}% Include figure files
\usepackage{dcolumn}% Align table columns on decimal point
\usepackage{bm}% bold math
\begin{document}

%\preprint{}

\title{Tunable thermal conductivity in defect engineered nanowires at low temperatures}


\author{Sajal Dhara}
\author{Hari S. Solanki}
\author{Arvind Pawan R.}
\author{Vibhor~Singh}
\author{Shamashis Sengupta}
\author{B.A. Chalke}
\affiliation{Department of Condensed Matter Physics and Materials Science, Tata Institute of Fundamental Research, Homi Bhabha Road,
Mumbai 400005, India}
\author{Abhishek Dhar}
\affiliation{Raman Research Institute, Bangalore, India  560080}
\author{Mahesh~Gokhale}
\author{Arnab~Bhattacharya}
\author{Mandar~M.~Deshmukh}
\email[]{deshmukh@tifr.res.in}

\affiliation{Department of Condensed Matter Physics and Materials Science, Tata Institute of Fundamental Research, Homi Bhabha Road,
Mumbai 400005, India}


\date{\today}

\begin{abstract}
We measure the thermal conductivity ($\kappa$) of individual InAs
nanowires (NWs), and find that it is 3 orders of magnitude smaller
than the bulk value in the temperature range of 10 to
50~K. We argue that the low $\kappa$ arises from the
strong localization of phonons in the random superlattice of
twin-defects oriented perpendicular to the axis of the NW. We
observe significant electronic contribution arising from the surface
accumulation layer which gives rise to tunability of $\kappa$ with the application
of electrostatic gate and magnetic field. Our devices
and measurements of $\kappa$ at different carrier
concentrations and magnetic field without introducing structural defects, offer a
means to study new aspects of nanoscale thermal transport.
\end{abstract}

\pacs{}
%\keywords{}

\maketitle

Thermal transport measurements on semiconducting nanowires (NWs) have attracted
a lot of attention in the last few years. Measurements of thermal
transport in nanostructures are important as they provide a platform
to test the existing descriptions of phonons in confined structures
and across complex interfaces \cite{complexthermoreview}, and have
the potential to result in technological applications as
thermoelectric systems
\cite{complexthermoreview,cahillreviewJAP,cahillreviewJHT}.
Different materials are benchmarked using the thermoelectric figure
of merit $ZT=\frac{S^2T}{\rho \kappa} $, where $S$ is the Seebeck
coefficient, $\rho$ is the electrical resistivity, $\kappa$ is the
thermal conductivity and $T$ is the absolute temperature. $ZT$ can be
increased by appropriate engineering of nanostructures. One of the
ways to increase $ZT$ is by reducing $\kappa$ without
 degrading its electrical conductivity and Seebeck coefficient \cite{hicks}. As a result,
 an ideal thermoelectric material
 is a glass for phonons and ordered for electronic transport.
Recently, several theoretical models as well as experimental studies
have been carried out in different semiconductor NWs like Si, Ge,
Bi$_2$Te$_3$ etc.
\cite{deyu,mingo2,renkun,hochbaum,jimheathsinanowire}. It is found
that for Si NWs, the value of $\kappa$ is reduced by two orders of
magnitude \cite{deyu,hochbaum,jimheathsinanowire} compared to bulk
values by tuning the roughness of the surface. III-V semiconductors
are also known to be good thermoelectric
materials, and theoretical studies suggest that InSb and InAs NWs
are good candidates for better $ZT$ \cite{mingo1}. InAs NWs have
been studied extensively to probe their charge and spin-transport
\cite{inas-spin-orbit-samuelson,our-inas,newpaper}. An aspect of InAs NWs
that makes studying their thermal transport, hitherto little
explored \cite{persson,shiinasnw}, of interest is the ability to
tune the density of twin defects and polytypes along its length by
varying growth parameters
\cite{twin-hiruma,xiong,caroffp,InP,petta-inas}. Exploiting this
control over crystal structure can help synthesize defect-engineered
NWs, whose lattice has aperiodic array of twins along its length
that modify phonon behavior, without significantly compromising
their electrical properties \cite{petta-inas}. Such NWs satisfy the
key criteria for a good thermoelectric material -- localization of
phonons without localizing electrons.

\begin{figure}
\includegraphics[width=85mm, bb=0 0 481.5 437]{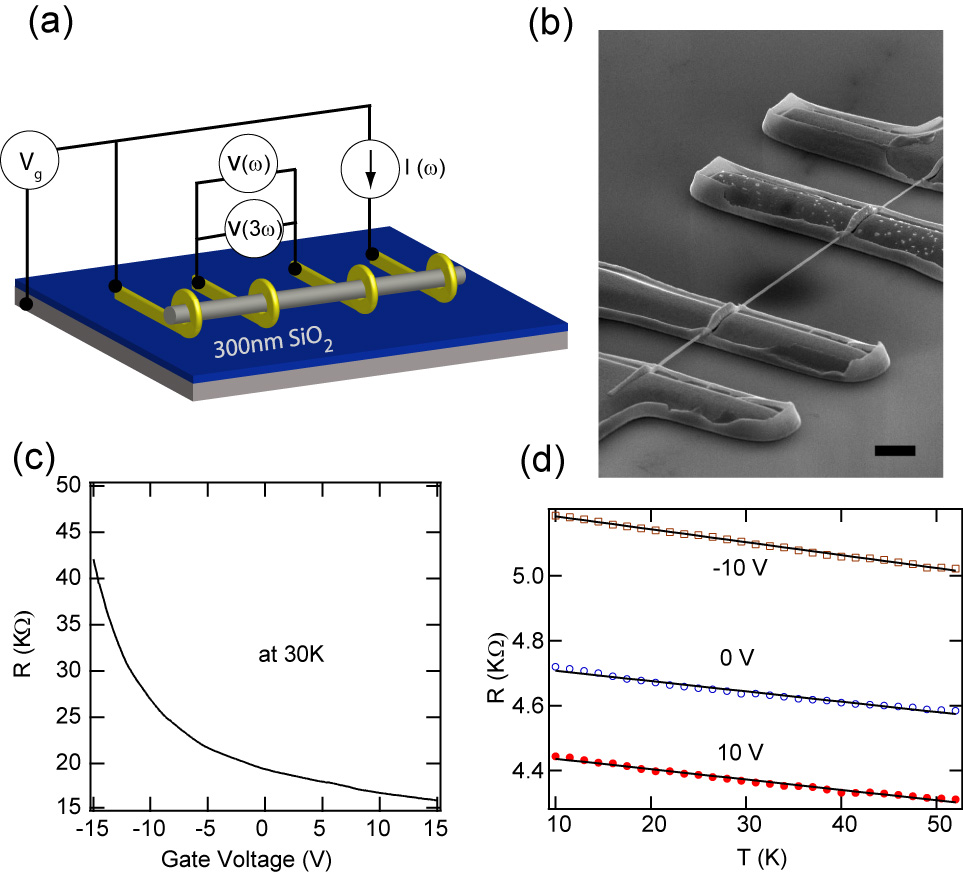}
\caption{ \label{fig:figure1} (color online) (a) Schematic of the four probe
suspended NW device and the circuit used for measuring
electrical signals for both electrical and thermal transport. (b) A tilted scanning electron microscope
image of a device. The scale bar corresponds to 1~$\mu$m. (c) Resistance as
a function of gate voltage V$_g$ at bath temperature of 30K for a typical device. (d) Resistance R as a function of
temperature for a device at three different gate voltages V$_g$. The solid curves
show linear fits to the data.}
\end{figure}

In this work we explore the $\kappa$ of suspended InAs NW field
effect transistors (FETs); the NWs have high density of aperiodic
twins and polytypes perpendicular to their axis
\cite{twin-hiruma,xiong,caroffp,petta-inas}. The random nature of
defects suppresses $\kappa_{ph}$ (the phonon contribution to
$\kappa$) resulting in reduction of thermal conductivity by 3 orders of magnitude from bulk.
Our method also has an advantage which is to study $\kappa$ as a function of carrier
concentration without introducing additional impurities. We observe significant change in the thermal conductivity with the application of gate voltage suggesting finite electronic contribution, and we propose that this contribution comes from the conduction electrons at the surface accumulation layer which is inherent to low band gap semiconductors like InAs and InN \cite{petrovykh,Tilburg,thickness}. To confirm the role of electronic contribution we apply magnetic field and see considerable changes in the thermal conductivity.

 The InAs NWs used to fabricate the devices in this work were
grown in a metal organic chemical vapour deposition (MOCVD) system
using the vapor-liquid-solid (VLS) technique on a  $< 111
>B$  oriented GaAs substrate. NWs of 100 - 200~nm diameter were used in this
experiment. 10-12~$\mu$m  long NWs were chosen for making four probe suspended devices on a $300$ nm thick,
SiO$_2$ coated degenerately doped Si wafer, which serves as the back
gate (growth and device fabrication details are given in supplementary S1 \cite{supp}). Fig.~\ref{fig:figure1}(a) shows the schematic of the device and the circuit for electrical and thermal measurements. Fig.~\ref{fig:figure1}(b)
shows a scanning electron microscope (SEM) image of a device. The
gold electrodes provide mechanical support and Ohmic contacts with negligible contact resistance to the
suspended NW besides thermally anchoring it at four points to the
bath temperature T$_b$. The electrical properties of the device are
characterized as a function of gate voltage (V$_g$) and temperature
before detailed thermal transport measurements.
Fig.~\ref{fig:figure1}(c) shows the resistance
(R=$\frac{V(\omega)}{I( \omega )}$) of a typical device as a function of
V$_g$ at a fixed value of T$_b$. The \emph{n}-type field effect
transistor (FET) behavior of the suspended NW devices is clearly
seen. Fig.~\ref{fig:figure1}(d) shows
the evolution of R as a function of temperature, T, and also the
corresponding linear fit in the temperature range 10-50 K.

\begin{figure}
\includegraphics[width=85mm, bb=0 0 478 446.5]{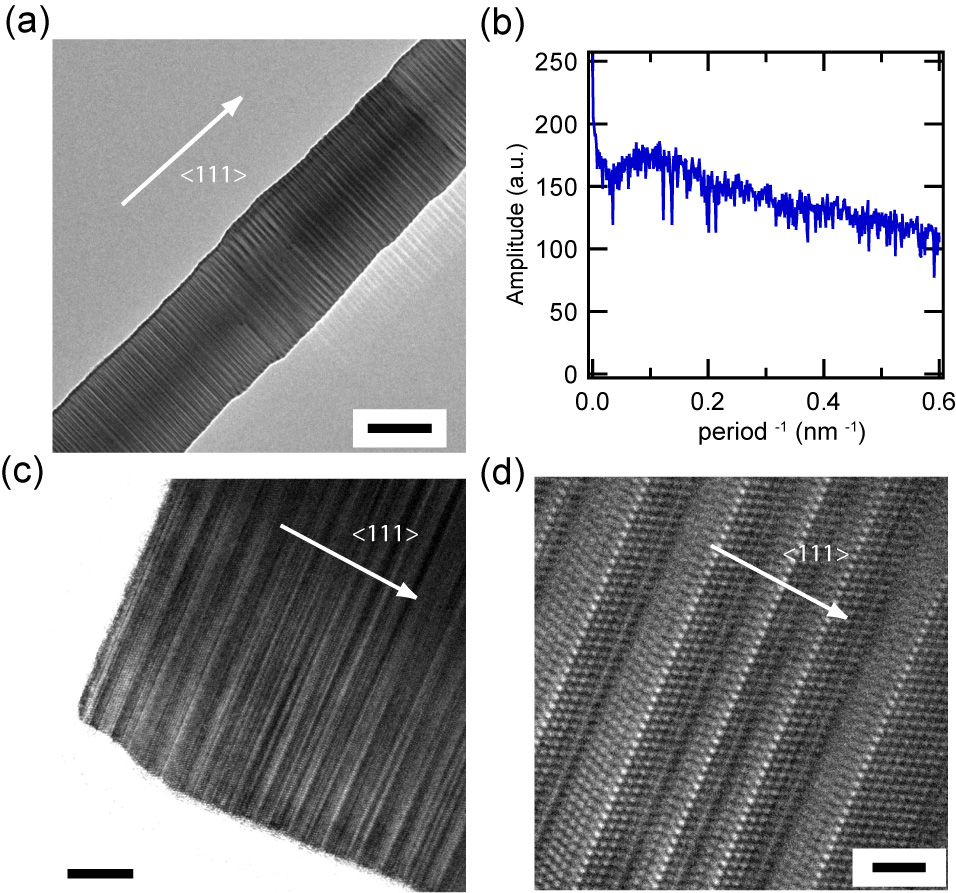}
\caption{ \label{fig:figure3} (color online) (a) TEM image of an
InAs NW shows twin boundaries perpendicular to the axis of the NW. The scale bar is 50~nm. (b) Fourier amplitude of the
intensity modulation along the length of the NW that shows no
sharp peaks. (c) A close-up
image showing the twin interfaces and the resulting zig-zag surface
roughness of the NWs. The scale bar is 10~nm. (d)
TEM image showing  the atomic fringes and the orientation of the
twin boundaries and polytypes. The scale bar is 2~nm. }
\end{figure}

High resolution transmission electron microscopy (TEM) studies were
used to characterize the microstructure of our twin-engineered NWs.
Fig.~\ref{fig:figure3}(a) shows a large scale TEM image of a NW
where we observe an aperiodic arrangement of the twin superlattice
perpendicular to its axis. The fact that the lattice has defects but
lacks a long-range periodicity is reflected in the Fourier amplitude
(seen in Fig.~\ref{fig:figure3}(b)) that has no sharp peaks along
the length of the NW. Figs.~\ref{fig:figure3}(c) \& (d) show a
detailed view of the twins and polytypes
\cite{twin-hiruma,xiong,caroffp,petta-inas}. The typical spacing of
these defects is $\sim$3-5~nm. Once electrical and structural
characterization is done, we perform simultaneous thermal and
electrical measurements.

\begin{figure}
\includegraphics[width=70mm, bb=0 0 440 624]{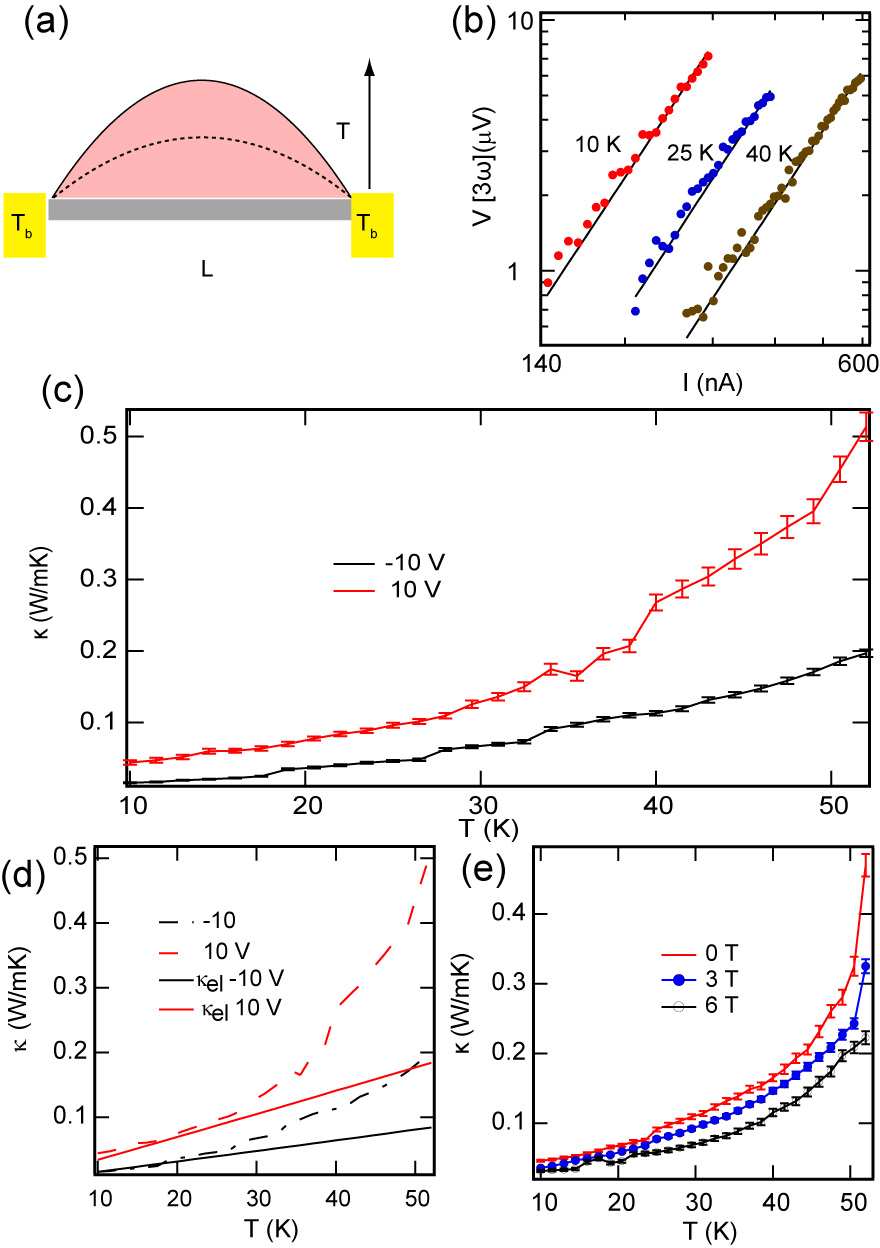}
\caption{ \label{fig:figure2} (color online) (a) The temperature
profile along the length L of the NW with the red shaded area
indicating the amplitude of oscillations. (b) The measured V(3$\omega$) signal as a function of
the excitation current I($ \omega $) for three different
temperatures T. The solid line shows the fit to the cubic power law
(Equation~\ref{eq:3omegaeqn}). (c) Plot of $\kappa$ as a function of temperature for two different gate
voltages. (d) Measured $\kappa$ as a function of temperature
(dotted lines) together with electronic contribution $\kappa_{el}$ (solid lines).
 (e) Measured $\kappa$ for the same device at
 three different magnetic fields
 as a function of the temperature at zero gate voltage.}
\end{figure}

One method of measuring thermal properties of
nanostructures uses a modification of the 3$\omega$ technique
\cite{cahill3omega} for $\kappa$ and specific heat measurement.
The modified 3$\omega$ technique
\cite{3omega,dames3omega,nickelnanowire} relies on fabricating
suspended four probe devices with the conducting NW serving both as
a heater and a thermometer.
The magnitude of V($3\omega$) can be related to $\kappa$ (in the limit
 $\omega \gamma \rightarrow 0 $ ) as \cite{3omega,dames3omega}
\begin{equation}
V(3 \omega)=\frac{4(I(\omega))^3 R R' L} {\pi^4\kappa A}, \label{eq:3omegaeqn}
\end{equation}
where $\omega$ is the angular frequency of the AC current,
$I(\omega)$ is its amplitude, $R$ and $R'$ are the resistance and
the derivative of resistance with respect to temperature
respectively, $L$ is the length of the NW between the two middle
probes and $A$, its cross section. \cite{3omega,dames3omega}. For
our devices, $L$ is $2-4 \mu$m and the thermal time constant
$\gamma$ can be estimated to be $\sim 10^{-5}$s (details of phase
information of $3 \omega$ signal in the limit of $\omega\gamma
\rightarrow 0$ are given in the supplementary S2 \cite{supp}).

To measure $\kappa$ at each temperature value, we vary the current
(I$(\omega)$) and obtain $\kappa$ from the fitting of
Equation~\ref{eq:3omegaeqn} to the measured V$(3 \omega)$ as a
function of injected I$(\omega)$. Fig.~\ref{fig:figure2}(b) shows
the evolution of V$(3 \omega)$ with I$(\omega)$ along with the cubic
power-law relationship that is essential to ensure that the $3
\omega$ technique is applicable within the applied current regime (to eliminate other source that can possibly show cubic scaling see supplementary S4 \cite{supp}).
As we vary the temperature, we also record the resistance. We
determine $R'$ from the derivative to the linear fit to the $R-T$ plot (see
Fig.~\ref{fig:figure1}(d)). The variation of $R$ with $T$ becomes small and non-monotonic above 50-60 K this is the reason we choose the temperature range of 10-50 K in our experiment. The phase of the $3 \omega$ signal is
measured to be zero at all temperatures (see supplementary Fig. S2 \cite{supp}). This procedure of fitting the power law given by
Equation~\ref{eq:3omegaeqn} at each instance of parameter space
allows us to extract $\kappa$ as a function of V$_g$ and
temperature. We have calibrated our technique using lithographically fabricated
170 nm wide 75 nm thick gold NWs in the temperature range of our measurement and obtained a Lorenz number around $2.4\times 10^{-8}$~$W\Omega~K^{-2}$ (see supplementary S3 \cite{supp}) which calibrates the measurement technique we have used.

Fig.~\ref{fig:figure2}(c) shows the plot of measured $\kappa$ as a
function of temperature at two values of V$_g$. The overall magnitude of $\kappa$ is about 1000
times smaller than the bulk values \cite{inasbulkkappa} (see
supplementary Fig. S14 \cite{supp}). This is a significant reduction in
$\kappa$ and the measured value is similar to the minimum thermal
conductivity model of Cahill \emph{et al.}
\cite{cahill-minimumthermal} (CWP model) based on the assumption
that the material is highly disordered. Similar reduction in $\kappa$,
beyond the CWP model, was also observed in disordered WSe$_2$ which
is a layered structure \cite{wse2}. Later we return to a model to
understand this better. Similar evolution of $\kappa$ is
also observed at different values of V$_g$ in all the measured
devices (see supplementary Fig. S 10-12 \cite{supp}).

Now we consider the effect of V$_g$ on $\kappa$. It is clear that
the electrical conductance of the semiconducting NW FET can be
modified as a function of V$_g$ (Fig.~\ref{fig:figure1}(c)).
However, the effect of electron density variation is not expected to
affect $\kappa$ in most semiconducting materials, as phonons are the
dominant carriers of heat \cite{zimanbook}. Direct calculation from electrical conductivity using the Wiedemann-Franz (WF) law gives a small value of the electronic contribution ($\kappa_{el}$) which does not explain the variation we see with V$_g$ (see supplementary S6 \cite{supp}). To understand the origin of large $\kappa_{el}$ we have to first understand the material property of the InAs wire. It is well known that in InAs there is an accumulation of charge at the surface. The resultant downward band bending pins the surface Fermi level above the conduction band minimum  \cite{petrovykh,thickness,Tilburg}. In a geometry where the surface to volume ratio is large, such as NWs, this surface accumulation charge (SAC) plays a major role in the electrical transport. The SAC takes part in the electrical conduction and scatters due to roughness at surface, which is also the reason for low electron mobility in InAs NWs compared to bulk \cite{our-inas,Tilburg}. We have passivated our devices using (NH$_4$)S$_x$ but this passivation leads to further enhancement to the surface charge density ($\sim10^{12}cm^{-2})$ \cite{petrovykh} and consequently excellent Ohmic contacts. Fig.~\ref{fig:figure1}(c) shows that our device is always in conducting state throughout the gate range with low resistance value, strongly suggesting the presence of the accumulation layer. To estimate the $\kappa_{el}$ we have to consider the thickness of the layer of this SAC since it will reduce the effective cross sectional area through which heat is transported by electrons. The electrons in the bulk and in the accumulation layer can be thought of two parallel channels conducting heat and we can define a effective layer thickness ($t$) to find out the actual $\kappa_{el}$ from the WF law,
\begin{equation}\kappa_{el}=\frac{l T_b L_0}{\pi R\{r^2-(r-t)^2\}}~~~(0<t\leq r)\end{equation}
where $l$ and $r$ are length and radius of the wire respectively, R is the electrical resistance at temperature $T_b$ and $L_0=2.45\times10^{-8}$~$W\Omega~K^{-2}$ is Lorenz number. We can get a rough estimate of $t$ from above equation by assuming that at lowest temperature the phonon contribution ($\kappa_{ph}$) is small and $\kappa_{el}\sim\kappa$. For the
NW of length 2.6 $\mu$m and radius 70 nm we get $t$ from our data as $\sim$20 nm and $\sim$10 nm for applied gate voltages of -10V and +10V respectively. Now we use Equation 2 to find $\kappa_{el}$ for all $T_b$, the result is plotted as solid lines in Fig.~\ref{fig:figure2}(d). This estimated $\kappa_{el}$ from a simplified picture captures the variation we see in our data with $V_g$.
Another confirmation about this non negligible $\kappa_{el}$ comes from our magnetic field measurement. One way to suppress the electronic contribution is to apply a magnetic field \cite{zimanbook,royal}, as it will reduce the electronic mean free path for heat transport by deflecting the semiclassical trajectory of electrons. We have done measurement in magnetic field up to 6T and see considerable reduction of thermal conductivity (see in Fig.~\ref{fig:figure2}(e)). Magnetic field can affect phonons via electron-phonon interaction, the amount of change in $\kappa_{ph}$ however, is much smaller \cite{elph} than the effect we see in our measurement.

\begin{figure}
\includegraphics[width=50mm, bb=0 0 420 392]{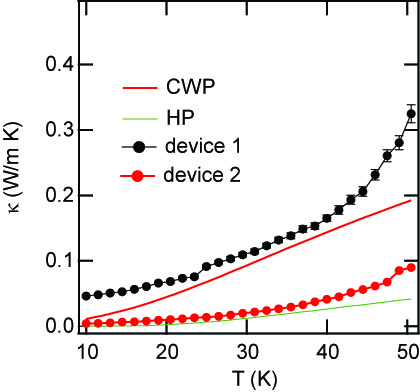}
\caption{ \label{fig:figure4} (color online) The plot of measured $\kappa$ as
a function of temperature of two different NW samples at 0 V$_g$, along with, the predicted values from two models to understand the reduction in
$\kappa_{ph}$ for InAs NWs. Based on CWP model of Cahill \emph{et al.}
\cite{cahill-minimumthermal}, and HP
model, of Hopkins \emph{et al.} \cite{minimumkappainterface}.}
\end{figure}

The total $\kappa$ is small and it is important to understand the
origin of reduction in $\kappa_{ph}$. It is clear from the
structural analysis in Fig.~\ref{fig:figure3} that our NWs have a
high density of twins. The twin defects introduce additional
interfaces and also cause the surface to be rough, with typical
roughness $\sim$2-3 nm. The heat conduction in bulk is dominated by
phonons whose mean free path $\lambda_{mfp}$ is much larger than the
spacing between defect planes throughout the temperature range of
our measurements (see supplementary Fig.~S16 \cite{supp}). The
average number of defect interfaces is $\sim$500 per $\mu$m -- this
makes the problem challenging to model comprehensively. However,
previous experiments on ordered
\cite{superlattice-venky,hurley-superlattice}, quasiperiodic
superlattices \cite{hurley-aperiodic} and calculations with
modulation of superlattice
\cite{mahan-minimumthermalconductivity,ren-superlattice} on similar
lengthscale~$<<\lambda_{mfp}$ suggest that suppression of $\kappa$
takes place in such structures. Our devices have aperiodic
interfaces
 (see Fig.~\ref{fig:figure3}(a)) and they are likely to attenuate a wide band of phonons. Additionally,
modeling indicates that heat transport across grain boundaries can
offer significant thermal impedance
\cite{maiti-acrossgrainboundaries}. In order to understand the
magnitude of the observed $\kappa$, we use a model by Hopkins
\emph{et al.} \cite{minimumkappainterface} (denoted by HP model)
which explicitly takes into account the role of interfaces for
calculating minimum $\kappa$ in layered materials.
We have calculated the predicted values of the model of Cahill \emph{et al. } \cite{cahill-minimumthermal} (denoted by CWP model) and the HP model for InAs (see supplementary S9 \cite{supp}). There is a range of $\kappa$ measured for devices made from different NW
samples (see supplementary S5 \cite{supp}); data from two such devices at 0~V$_g$ is shown in Fig.~\ref{fig:figure4}. Without introducing any
free parameters, the predictions of the CWP and HP models result in values that are comparable to the range in $\kappa$ measured in our experiments.
%As expected the estimates of the interface-scattering inclusive HP model are lower than that of the CWP model.

In summary, we demonstrate that thermal conductivity of InAs NWs
with twin defects and polytypes is 1000 times smaller than bulk
values which will make our devices suitable candidates for
good thermoelectric materials. We show the electronic contribution
becoming significant due to presence of the surface charge accumulation.
The ability to tune thermal conductivity electrostatically can lead to
novel ways of probing heat transport in defect engineered nanostructures,
and possible device applications.

We would like to acknowledge the Government of India for support
(TIFR grant numbers 11P803, 11P809 and 11P812).

\end{document}